\documentclass[letterpaper, 10 pt, conference]{ieeeconf}  
\IEEEoverridecommandlockouts                              
\usepackage{cite}
\usepackage{amsmath,amssymb,amsfonts}
\usepackage{multirow}
\usepackage{ctable}
\usepackage[linesnumbered,ruled,vlined]{algorithm2e}
\usepackage{graphicx}
\usepackage{textcomp}
\usepackage{xcolor}
\def\BibTeX{{\rm B\kern-.05em{\sc i\kern-.025em b}\kern-.08em
		T\kern-.1667em\lower.7ex\hbox{E}\kern-.125emX}}

\usepackage{bm}

\begin{document}
	
	\title{\LARGE \bf
		Gaussian Process-driven Hidden Markov Models for Early Diagnosis of Infant Gait Anomalies}

	\author{Luis F. Torres-Torres, Jonatan Arias-Garc\'ia, Hern\'an F. Garc\'ia, \\ Andr\'es F. L\'opez-Lopera, and Jesús F. Vargas-Bonilla
		\thanks{*This research work was conducted under the project COL111089784907 funded by MINCIENCIAS (Colombia).}
		\thanks{
			Luis F. Torres-Torres, Hern\'an F. Garc\'ia, and Jesús F. Vargas-Bonilla are with the SISTEMIC Research Group, Universidad de Antioquia, Medellín, Colombia, {\small \{luis.torres1, hernanf.garcia, jesus.vargas\}@udea.edu.co}. 
			Jonatan Arias-Garc\'ia is with the Automatics Research Group, Universidad Tecnologica de Pereira, Pereira, Colombia, {\small jonnatan.arias@utp.edu.co}.
			Andr\'es F. L\'opez-Lopera is with CERAMATHS, Universit\'e Polytechnique Hauts-de-France, Valenciennes, France, {\small andres.lopezlopera@uphf.fr}.}%
	}

	\maketitle
	\thispagestyle{empty}
	\pagestyle{empty}
	
	\begin{abstract}
		Gait analysis is critical in the early detection and intervention of motor neurological disorders in infants. Despite its importance, traditional methods often struggle to model the high variability and rapid developmental changes inherent to infant gait. To address these challenges, we propose a probabilistic Gaussian Process (GP)-driven Hidden Markov Model (HMM) to capture the complex temporal dynamics of infant gait cycles and enable automatic recognition of gait anomalies. We use a Multi-Output GP (MoGP) framework to model interdependencies between multiple gait signals, with a composite kernel designed to account for smooth, non-smooth, and periodic behaviors exhibited in gait cycles. The HMM segments gait phases into normal and abnormal states, facilitating the precise identification of pathological movement patterns in stance and swing phases. The proposed model is trained and assessed using a dataset of infants with and without motor neurological disorders via leave-one-subject-out cross-validation. Results demonstrate that the MoGP outperforms Long Short-Term Memory (LSTM) based neural networks in modeling gait dynamics, offering improved accuracy, variance explanation, and temporal alignment.
		Further, the predictive performance of MoGP provides a principled framework for uncertainty quantification, allowing confidence estimation in gait trajectory predictions. Additionally, the HMM enhances interpretability by explicitly modeling gait phase transitions, improving the detection of subtle anomalies across multiple gait cycles. These findings highlight the MoGP-HMM framework as a robust automatic gait analysis tool, allowing early diagnosis and intervention strategies for infants with neurological motor disorders.
		
	\end{abstract}


	\section{Introduction}
	
	Gait analysis is a fundamental tool for evaluating and diagnosing motor neurological disorders, particularly in pediatric populations \cite{whittle2014gait}. Accurate evaluation of infant gait patterns can provide critical insights into neuromotor development and facilitate early intervention strategies \cite{williams2009gait}. With the advancement of motion capture technologies and the integration of artificial intelligence (AI) in healthcare, there is a growing interest in developing automated systems to recognize abnormal gait patterns \cite{chambers2018review}.
	
	Previous research has extensively explored various techniques for automatic gait analysis, including traditional statistical methods, machine learning algorithms, and biomechanical modeling. In particular, machine learning approaches such as Support Vector Machines (SVMs) \cite{begg2005support}, Hidden Markov Models (HMMs) \cite{mantyjarvi2005identifying}, and Neural Networks (NNs) \cite{jin2007neural} have been employed to classify and predict gait abnormalities. Deep learning models, particularly Convolutional Neural Networks (CNNs), have shown significant promise in capturing spatial-temporal features of gait data \cite{wan2018survey,zeng2016convolutional}. Additionally, studies have used Gaussian Processes (GPs) to model gait dynamics due to their probabilistic nature and flexibility \cite{romer2012probabilistic,dai2017gaussian}.
	
	Despite these advancements, challenges remain in modeling infant gait data's complex temporal dynamics and interdependencies. Existing models often struggle with the high variability and rapid developmental changes characteristic of infant gait \cite{dompseler2019variability}. Moreover, many approaches focus on single-output predictions and do not adequately capture the correlations between multiple gait parameters \cite{begg2005support}. This limitation underscores the need for more sophisticated modeling techniques capable of handling multi-dimensional outputs and accounting for intricate relationships between different aspects of movement.
	
	This study proposes a novel application of Multi-Output Gaussian Processes (MoGP) to model and recognize abnormal gait patterns in infants with motor neurological disorders. Using their flexibility and probabilistic framework, we aim to capture gait data's temporal dynamics and interrelated signals effectively. Our approach involves extracting key gait signals, constructing a composite kernel to model complex patterns, and implementing a multitasking framework to account for inter-signal correlations.
	
	The proposed method offers several advantages. It quantifies uncertainty through probabilistic predictions, accommodates multiple outputs simultaneously, and can model nonlinear and nonstationary patterns prevalent in gait data. By addressing the limitations of previous studies, our work contributes to an advancement in the field of automatic gait analysis. It can potentially improve early diagnosis and intervention strategies for infants with neurological motor disorders, ultimately enhancing patient outcomes \cite{chen2015recent,wang2019deep}.
	
	Our main contributions are as follows:
	\begin{itemize}
		\item We introduce a MoGP framework for modeling multi-dimensional infant gait data, capturing temporal dynamics and interdependencies between gait parameters.
		\item We design a composite kernel combining different kernels to model complex gait patterns accurately.
		\item We validate our approach through experiments on real-world infant gait databases, demonstrating improved performance over traditional single-output models.
	\end{itemize}
	
	The rest of this paper proceeds as follows. Section \ref{sec:Methodology} presents the proposed MoGP framework, including the composite kernel design and multitask formulation. Section \ref{sec:results} discusses the results, highlighting the model's performance in capturing gait dynamics and recognizing anomalies. Finally, Section \ref{sec:conclusions} concludes the paper by summarizing key findings, discussing clinical implications, and outlining future research directions.
	
	\section{Methodology}
	\label{sec:Methodology}
	
	\subsection{Data Acquisition and Pre-processing}
	\label{sec:Methodology:subsec:database}

	Kinematic gait data were collected from 80 participants, including 59 infants with motor neurological disorders (e.g., Cerebral Palsy, Down Syndrome, Autism, and rare diseases) and 21 typically developing infants as a control group. The data consists of three-dimensional displacement vectors representing joint movements during gait cycles, recorded at $T$ time points per cycle with a motion capture system sampling at 30 FPS. The dataset is denoted as $\mathbf{X} \in \mathbb{R}^{N \times T \times D}$, where $N = 80$ is the number of infants, $T = 400$ represents time points, and $D = 3$ denotes spatial dimensions. We used synchronized cameras, OpenPose \cite{OpenPose} estimated 2D joint positions, which were triangulated with Pose2Sim \cite{Pose2Sim} to obtain 3D joint trajectories. Moreover, data were preprocessed with a low-pass Butterworth filter to reduce noise, statistical imputation to correct joint estimations, normalization to minimize inter-subject variability, and temporal alignment to account for differing gait cycle durations.
	
	This study focused on the $y$-axis (vertical displacement) due to its clinical relevance in detecting gait abnormalities, such as altered knee lifting or hip displacement, which are key indicators of pathological gait. Analyzing vertical dynamics highlights these features while minimizing less significant lateral and forward. The six gait signals are then concatenated to form a vector-valued function:
	\begin{equation*}
		\bm{y}(t) = 
		\left[ 
		{y}_{\text{H}}^{(r)}(t), \;
		{y}_{\text{H}}^{(l)}(t), \;  
		{y}_{\text{K}}^{(r)}(t), \;
		{y}_{\text{K}}^{(l)}(t), \;  
		{y}_{\text{A}}^{(r)}(t), \;
		{y}_{\text{A}}^{(l)}(t)
		\right]^{\top},
	\end{equation*}
	where $t \in [0, 1]$ is the (normalized) time within the gait cycles, and $y_{a}^{b}(t)$, with $a \in \{\text{H} := \text{Hip}, \text{K} := \text{Knee}, \text{A} := \text{Ankle}\}$ and $b \in \{r := \text{right}, l := \text{left}\}$, is the $y$-coordinate gait signal of the joint $a$ viewed from the $b$ side. 
	
	\subsection{Multi-output Gaussian Process (MoGP) Model}
	\label{sec:Methodology:subsec:MoGP}
	We consider an MoGP to capture the temporal dynamics and interdependencies of the gait signals. Loosely speaking, it defines a GP prior over a vector-valued function $\bm{y}(t)$:
	\[
	\bm{y}(\cdot) = 
	\left[ 
	{y}_{1}(\cdot), \ldots, {y}_{M}(\cdot)
	\right]^{\top}
	\sim \mathcal{GP}(\mu(\cdot), k(\cdot, \cdot)),
	\]
	where $y_m(\cdot)$ is the scalar-valued function of the $m$-th output for $m \in \{1, \ldots, M\}$ with $M \in \mathbb{N}$ denoting the number of outputs. In our case, $M = 6$. The $\mu(\cdot)$ and $k(\cdot, \cdot)$ are the mean and kernel functions of the GP prior. 
	
\begin{figure*}[t!]
		\centering
		\includegraphics[height=0.2\linewidth]{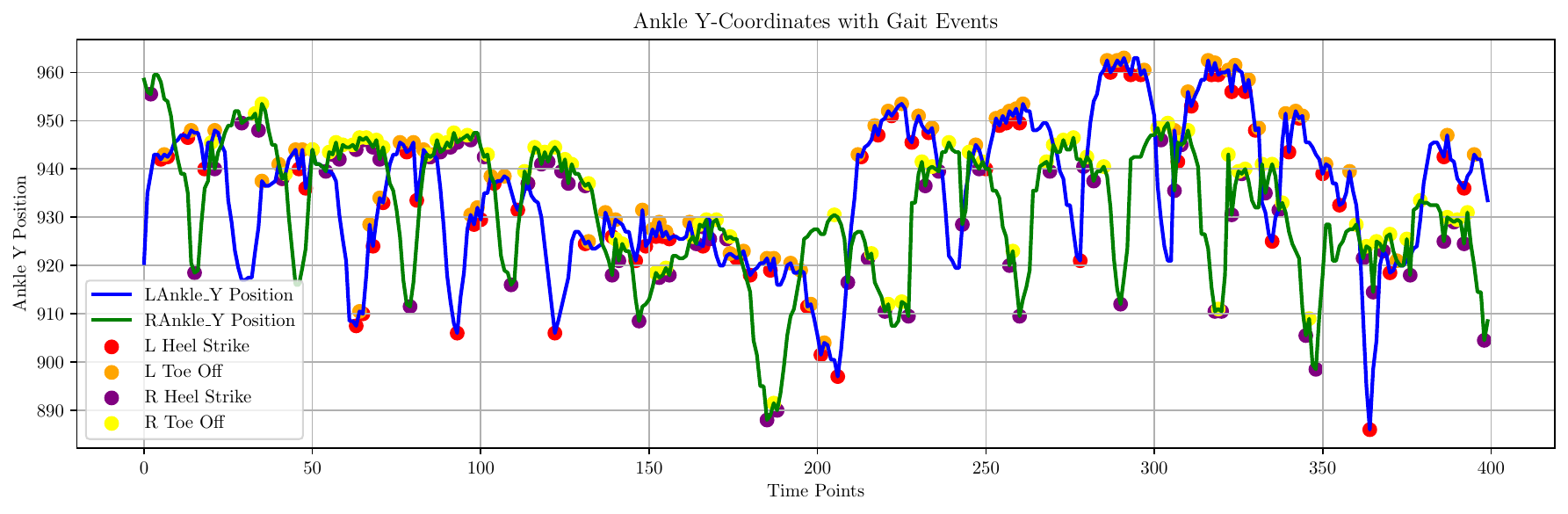}
		\includegraphics[height=0.2\linewidth]{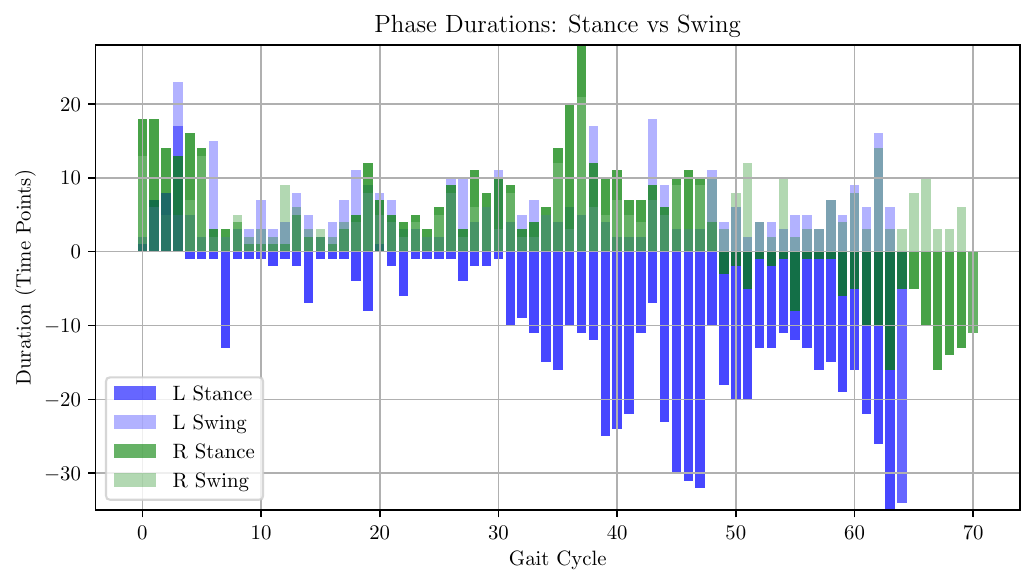}

		\caption{Visualization of Gait Events, Joint Angles, and Phase Durations. The top plots show the $y$-coordinates of the left and right ankles with markers for heel strikes and toe-offs, highlighting key gait events. The middle plots show the knee angles for both legs, illustrating the flexion and extension throughout the gait cycle. The bottom plot presents the phase durations of stance and swing phases for the left and right legs, showing consistency and symmetry across multiple gait cycles. These visualizations provide a comprehensive view of gait dynamics, supporting the identification of potential gait abnormalities and aiding clinical assessments.}
		\label{fig:PhaseDuration}
	\end{figure*}

	In practice, the individual terms in the mean function $\mu(\cdot) = (\mu_m(\cdot))_{1 \leq m \leq M}$ are typically assumed to be constant parameters that can be estimated within the MoGP framework. Hence, the kernel function $k(\cdot, \cdot)$ is crucial in defining prior beliefs about the target functions. In this study, we adopt the multi-output kernel introduced in~\cite{Bonilla2007MTGP}, which is based on the intrinsic model of coregionalisation (ICM):
	\begin{equation}
		k_{m, m'}(t, t')
		:= \operatorname{cov}\left( y_m(t), y_{m'}(t') \right)
		= B_{m, m'} \cdot k_t(t, t'),
	\end{equation}
	where $\bm{B} = (B_{m,m'})_{1 \leq m, m' \leq M}$ is a coregionalisation matrix that defines the correlations between different outputs. We design the kernel $k_t(\cdot, \cdot)$ based on expert knowledge to account for the gait behaviors observed in Figure~\ref{fig:PhaseDuration}. We adopt the following customized composite structure:
	\begin{equation}
		k_t(t, t') =  k_{\text{Per}}(t, t') 
		+ k_{\text{SE}}(t, t')
		+  k_{\text{Mat}}(t, t'),
		\label{eq:compositek}
	\end{equation}
	where $k_{\text{Per}}(\cdot, \cdot)$ is a periodic kernel, $k_{\text{SE}}(\cdot, \cdot)$ is the Squared Exponential (SE) kernel, and $k_{\text{Mat}}(\cdot, \cdot)$ is the Mat\'ern 3/2 kernel. Next, we present the expressions for these three sub-kernels and justify their selection.
	
	
	\smallskip
	
	\textbf{Periodic kernel:} This kernel is used to capture the periodic behaviors in the gait cycles, and is given by
	\[
	k_{\text{Per}}(t, t') = \sigma^2_{\text{Per}} \exp \Bigg( -\frac{2\sin^2\left( \frac{\pi |t - t'|}{p} \right)}{\ell_{\text{Per}}^2} \Bigg),
	\]
	with $p > 0$ the period length parameter.
	
	\smallskip
	
	\textbf{SE kernel:} This kernel is considered to model smooth global gait variations. It is the most widely used in GP frameworks and is given by:
	\[
	k_{\text{SE}}(t, t') = \sigma^2_{\text{SE}} \exp\left( -\frac{(t - t')^2}{2\ell_{\text{SE}}^2} \right).
	\]
	
	\smallskip
	
	\textbf{Matérn 3/2 kernel:} It targets local non-smooth variations in the gait cycles, and is given by
	\[
	k_{\text{Mat}}(t, t') = \sigma_{\text{Mat}}^2 \left(1 + \sqrt{3} \frac{|t-t'|}{\ell_{\text{Mat}}}\right) \exp\left( -\sqrt{3} \frac{|t - t'|}{\ell_{\text{Mat}}} \right).
	\]

		
		
	
	\medskip
	
	The parameters $(\sigma_{\text{Per}}^2, \sigma_{\text{SE}}^2, \sigma_{\text{Mat}}^2)$ and $(\ell_{\text{Per}}, \ell_{\text{SE}}, \ell_{\text{Mat}})$ are the variances and the length-scales, respectively, of the sub-kernels discussed above. Along with the coregionalisation coefficients in $\bm{B}$, they are optimized via maximum likelihood~\cite{Bonilla2007MTGP}, using the gait signals from all subjects.

	\subsection{Anomaly Detection via Hidden Markov Model (HMM)}
	\label{sec:Methodology:subsec:HMM}
	
	To detect abnormal gait patterns, we applied an HMM to each Ankle joint signal, $\bm{o}^{(t)} = [ y_{\text{A}}^{(r)}(t), \ y_{\text{A}}^{(l)}(t) ]^\top$, predicted by the MoGP. Note that the superscript notation is used here because functions are treated as discrete signals in HMM models. Considering both signals allows for the capture of bilateral gait dynamics. The HMMs here targets the probabilistic modeling of the gait phases:
	\begin{align*}
		S 
		= \{ s_1, s_2, s_3, s_4 \}
		= \{ 
		& \text{normal stance}, \text{normal swing},
		\\
		& \text{abnormal stance}, \text{abnormal swing} \}. 
	\end{align*}
	Hence, we consider HMMs with $|S| = 4$ hidden states.
	
	\smallskip
	
	\subsubsection{Gait Event Detection}
	Heel strike and toe-off events are detected from the MoGP predictions of the Ankle joints:
	\begin{align*}
		\text{Heel Strikes} &: \left\{ t_i \mid y_{\text{A}}(t_i) \text{ is a local minimum} \right\}, \\
		\text{Toe-Offs} &: \left\{ t_i \mid y_{\text{A}}(t_i) \text{ is a local maximum} \right\}.
	\end{align*}
	These events can be determined numerically for both signals in $\bm{o}^{(t)}$ as the latter are predicted using the MoGP for all $t \in [0, 1]$. We apply signal-processing techniques to identify these events.
	
	\smallskip
	
	\subsubsection{HMM settings} 
	\begin{itemize}
		\item \textbf{Initial state probabilities $\bm{\pi}$.} To favor the normal stance, 
		we set $\bm{\pi} = [0.6, \; 0.3, \; 0.05, \; 0.05]^\top$. This reflects a higher likelihood of the gait cycle beginning in the phase $s_1$. 
		
		\item \textbf{State transition probabilities $\bm{A}$.} 
		We design $\bm{A}$ based on expert knowledge of gait phase transitions:
		\begin{equation*}
			\bm{A} = \begin{bmatrix}
				0.7 & 0.25 & 0.05 & 0.0 \\
				0.3 & 0.6 & 0.0 & 0.1 \\
				0.25 & 0.2 & 0.5 & 0.05 \\
				0.25 & 0.2 & 0.05 & 0.5 \\
			\end{bmatrix}.
		\end{equation*}
		It assigns higher (respectively, lower) probabilities to transitions within normal (resp. into abnormal) phases.
		
		\item \textbf{Emission probabilities $P(\bm{o}^{(t)} \mid s^{(t)} = s_i)$.} 
		As assumed in the HMM literature, we consider observation likelihoods $p(\bm{o}^{(t)} \mid s^{(t)} = s_i) = \mathcal{N}(\bm{\mu}_i, \bm{\Sigma})$, where $\bm{\mu}_i$ is the mean observation vector for state $s_i$, and $\bm{\Sigma}$ is the covariance matrix that is shared across different states.
	\end{itemize}
	
	\smallskip
	
	\subsubsection{Model Training and Inference}
	The HMM parameters $\bm{\theta} = (\bm{\pi}, \bm{A}, (\bm{\mu}_i)_{1 \leq i \leq 4}, \bm{\Sigma})$ are estimated via Baum-Welch Expectation-Maximization (EM) algorithm which seeks to maximize the likelihood of the observations:
	\begin{equation*}
		\bm{\theta}_{\star} = \underset{\bm{\theta}}{\arg\max} \ p(\bm{O} \mid \bm{\theta}),
	\end{equation*}
	where $\bm{O} = (\bm{o}^{(1)}, \dots, \bm{o}^{(T)})$ is the sequence of observations.
	
	For inference, the Viterbi algorithm is used to compute the most probable sequence of the hidden states $\widehat{s}^{(1)}, \dots, \widehat{s}^{(T)}$ given the observations and the estimated model parameters:
	\begin{equation*}
		(\widehat{s}^{(1)}, \dots, \widehat{s}^{(T)} ) 
		= \underset{{s}^{(1)}, \dots, {s}^{(T)}}{\arg\max} \ p\left( {s}^{(1)}, \dots, {s}^{(T)} \mid \bm{O}, \bm{\theta}_{\star} \right).
	\end{equation*}
	
	\smallskip
	
	\subsubsection{Anomaly Detection Criteria}
	Anomalies are identified by analyzing the estimated hidden states. Time instants classified as either ``abnormal stance'' ($s_3$) or ``abnormal swing'' ($s_4$) are marked as anomalous segments:
	\begin{equation*}
		\text{Anomalous Segments} : \left\{ t \mid \widehat{s}^{(t)} \in \{ s_3, s_4 \} \right\}.
	\end{equation*}
	This approach enables the temporal localization of gait abnormalities, supporting detailed clinical assessments.
	
	\subsection{Numerical Implementation}
	\label{sec:Methodology:subsec:implementations}
	The multioutput Gaussian process (MoGP) and the hidden Markov model (HMM) were implemented in PyTorch \cite{Paszke2019Pytorch}, using the \texttt{ gptorch} \cite{Jacob2018GPtorch} and \texttt{hmmlearn} \cite{hmmlearn} libraries, respectively. Model optimization was performed using the Adam algorithm \cite{Kingma2015Adam}, with a learning rate of $7.5 \times 10^{-3}$ to ensure efficient convergence while mitigating the risk of early stagnation. A weight decay regularization term was introduced to control model complexity and prevent overfitting, particularly when handling high-variance gait data.
	
	The MoGP framework was configured with a composite kernel structure, incorporating Matérn, Radial Basis Function (RBF), and periodic components to capture smooth, non-smooth, and cyclic variations in gait trajectories. The covariance structure was formulated within a multitask framework to model interdependencies among joint movements. The inference process was performed using exact Gaussian process regression, ensuring analytically tractable posterior distributions.
	
	For temporal segmentation, the HMM was designed with four latent states corresponding to distinct gait phases. The emission probabilities were modeled using Gaussian distributions parameterized by the MoGP-predicted ankle trajectories. The Baum-Welch algorithm was employed for parameter estimation, while the Viterbi algorithm was used for optimal state sequence inference.

	

	\section{Results}
	\label{sec:results}
	
	
	
	
	\begin{figure*}[ht]
		\centering
		
		\includegraphics[width=0.49\linewidth]{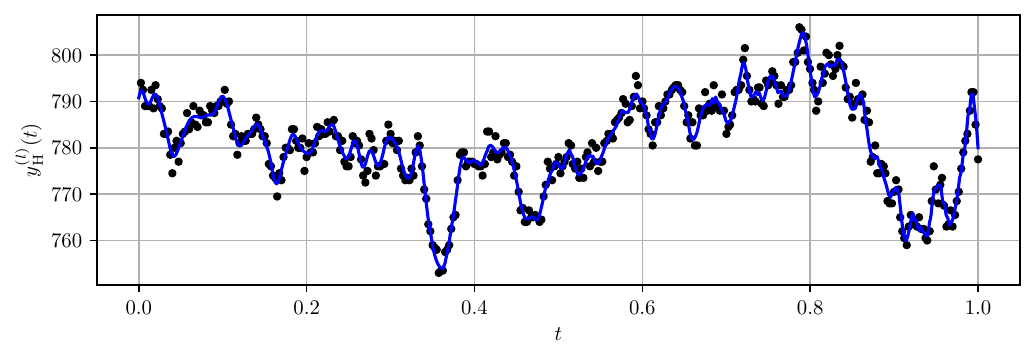}    
		\includegraphics[width=0.49\linewidth]{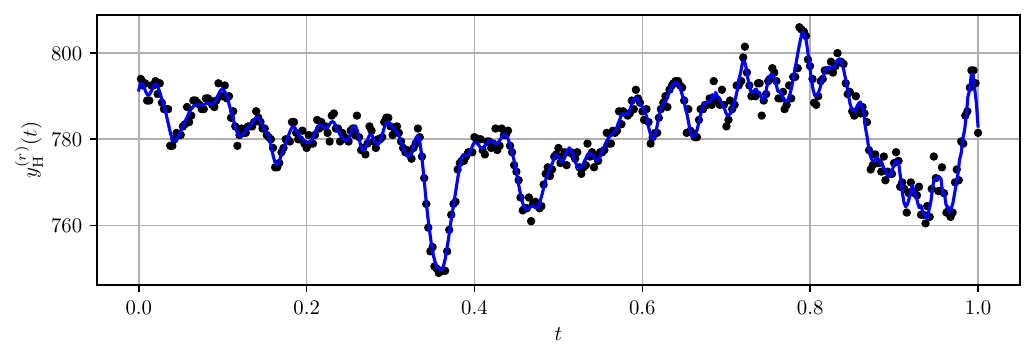}        
		\vspace{-4.2ex}    
		
		\includegraphics[width=0.49\linewidth]{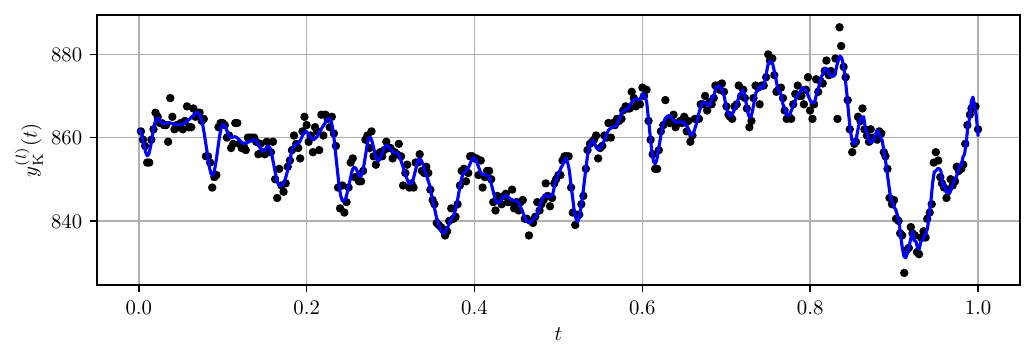}    
		\includegraphics[width=0.49\linewidth]{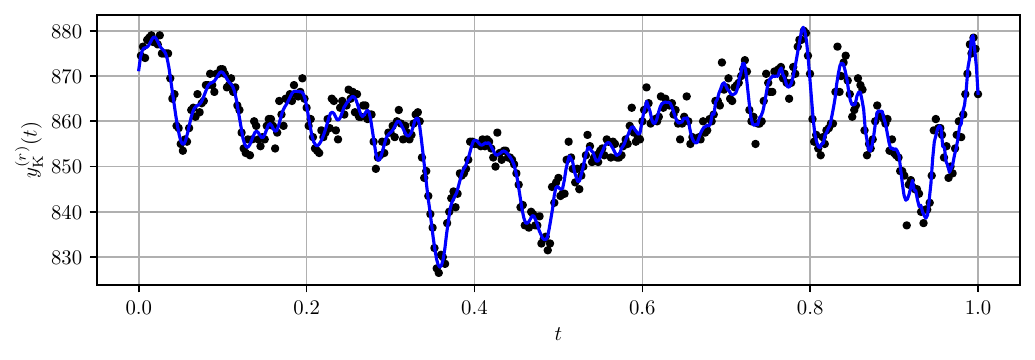}
		\vspace{-4.2ex}    
		
		\includegraphics[width=0.49\linewidth]{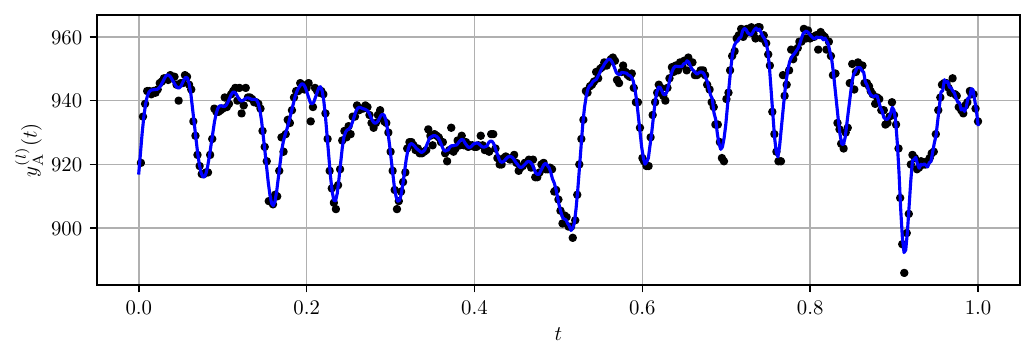}
		\includegraphics[width=0.49\linewidth]{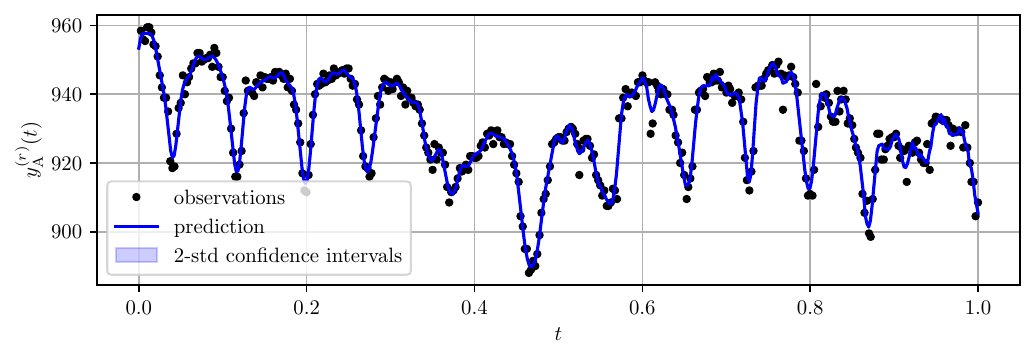}   
		\vspace{0ex}
		
		\caption{Predicted $y$-coordinate gait trajectories for the hip, knee, and ankle joints (top to bottom) on both the left and right sides of the body (left to right) using the MoGP model. The panels show the observations (black markers), and the predictions (solid lines) led by the MoGP model, together with the 2-standard deviation predictive intervals (shaded regions).}
		\label{fig:Predictions}
	\end{figure*}
	

	\subsection{MoGP predictions}
	
	The results presented in Figure \ref{fig:Predictions} illustrate the predictive capabilities of the MoGP model in capturing joint trajectories throughout the gait cycle. The model accurately aligns observed and predicted joint displacements across all six lower limb key points with well-calibrated predictive intervals.   The shaded confidence bands highlight the uncertainty associated with the predictions, effectively predicting regions where deviations from the expected gait trajectory occur. This is particularly valuable for detecting abnormal gait patterns, as areas of increased uncertainty may correspond to irregular joint motion. 
	
	The performance of the proposed MoGP framework stems from the model's use of three sub-kernels, as discussed in Section \ref{sec:Methodology:subsec:MoGP}. The Matérn 3/2 kernel captures non-smooth transitions and irregular movements, the SE kernel models continuous variations, and the periodic kernel captures the cyclic patterns inherent in gait cycles. The performance with individual kernels is also reported in Table \ref{tab:results_comparison}, showing that the MoGP with our customized kernel function in \eqref{eq:compositek} outperforms the other configurations.

	\begin{table*}[t!]
		\centering
		\caption{Performance comparison between MoGP and LSTM considering different settings. The best results are highlighted in bold. 
		}
		\label{tab:results_comparison}
			\begin{tabular}{c|ccc|cccc}
				\toprule
				\multirow{2}{*}{\textbf{Metric}} & \multicolumn{3}{c|}{\textbf{MoGP}} & \multicolumn{4}{c}{\textbf{LSTM}} \\
				& $k_{\text{SE}}$ & $k_{\text{Per}}$ 
				&  $k_{\text{SE}} + k_{\text{Per}} + k_{\text{Mat}}$ 
				& { $L = 2$, $H=128$} & { $L = 5$, $H=50$} & { $L = 10$, $H=50$} & { $L = 10$, $H=128$}\\
				\midrule
				MAE             & 2.875 & 5.894  
				& \textbf{1.659} 
				& 5.794 & 6.056 & 5.984 & 5.950\\
				
				$R^2$           & 0.895 & 0.583 
				& \textbf{0.963} 
				& 0.591 & 0.553  & 0.574 & 0.592 \\
				aDTW      & 56.982 & 115.678 
				& \textbf{34.920} 
				& 86.995 & 95.667 & 90.097 & 84.306\\
				
				\bottomrule
			\end{tabular}
	\end{table*}

\begin{figure}[t!]
		\centering
		\includegraphics[width=0.85\linewidth]{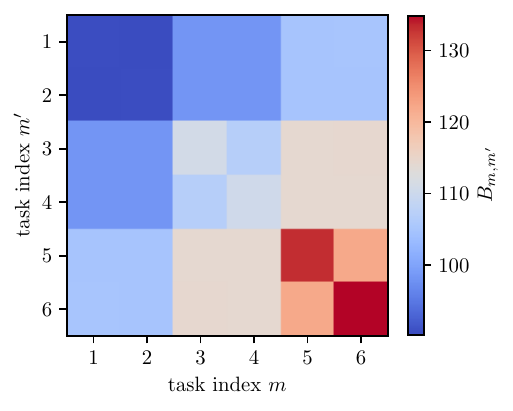}
		\caption{Coregionalisation matrix $\bm{B} = (B_{m,m'})_{1 \leq m, m' \leq 6}$ with outputs sorted as follows:  
			$m \in \{
			1:\text{Hip-Right}, \; 2:\text{Hip-Left},
			3:\text{Knee-Right}, \; 4:\text{Knee-Left},
			5:\text{Ankle-Right}, \; 6:\text{Ankle-Left} \}$. A higher value corresponds to a higher correlation between the outputs.
		}
		\label{fig:CorMatrix}
	\end{figure}

	
	The coregionalisation matrix $\bm{B}$ enables information sharing across six output tasks, corresponding to key joints such as the hips, knees, and ankles. This structure enhances prediction consistency by capturing correlations between joint movements. For instance, the model effectively reflects biomechanical dependencies, such as how hip motion influences ankle movement during different gait phases. Figure~\ref{fig:CorMatrix} shows that the highest correlations are observed between bilateral joint pairs, particularly at the knee and ankle levels, reflecting the inherent symmetry of human gait. Moderate correlations between the hip and knee indicate coordinated movement within the lower limb. However, lower correlations between the hip and ankle suggest greater independence in their motion, as expected from biomechanical constraints. These findings validate MoGP’s ability to capture structured dependencies within gait kinematics, reinforcing its suitability for clinical gait analysis and robotic locomotion applications. 
	
	\subsection{Numerical Comparison with LSTM Models}
	
	The performance of the proposed MoGP model is compared with a Long Short-Term Memory (LSTM)-based neural network using five evaluation metrics: Mean Absolute Error (MAE), Coefficient of Determination ($R^2$) \cite{devore2011probability}, and average Dynamic Time Warping (aDTW) distance \cite{Petitjean2011DTW, liu2019adaptive}. 
	
	
	
	
	Table~\ref{tab:results_comparison} presents the performance metrics of the MoGP compared to LSTM models across different configurations. MoGP with the customized kernel demonstrates superior predictive accuracy, achieving an MAE of 1.659, substantially lower than the best-performing LSTM configuration, which obtained an MAE of 5.794. This implies that MoGP effectively minimizes prediction errors, providing a more precise estimation of gait trajectories compared to LSTM.
	
	The $R^2$ score further highlights MoGP's advantage, reaching 0.963 versus 0.591 for LSTM, indicating that MoGP captures a significantly larger proportion of variance in gait data. This results in a more robust and reliable representation of movement dynamics, whereas the lower $R^2$ score of LSTM suggests limitations in generalization, likely due to the complexity of long-term dependencies in gait sequences and the challenge of training deep models with limited data.
	
	The aDTW distance evaluates the temporal alignment, where lower values indicate better alignment between predicted and actual trajectories. MoGP achieves an aDTW of 34.920, which is 59.8\% lower than the best LSTM result (86.995), confirming MoGP's superior ability to preserve the temporal structure of gait patterns. This is particularly relevant in clinical gait analysis, where precise trajectory alignment is crucial for detecting subtle abnormalities and assessing rehabilitation progress.
	
	Overall, the results in Table~\ref{tab:results_comparison} demonstrate that MoGP consistently outperforms LSTM in predictive accuracy, variance explanation, and temporal alignment. These findings underscore MoGP's potential for robust gait modeling and abnormal movement detection in biomedical applications.
	
	\subsection{Analysis of gait events}
	
	Figure \ref{fig:PhaseDuration} illustrates key aspects of gait dynamics, including ankle positions, knee angles, and phase durations for stance and swing cycles across time points and gait cycles. The plotted ankle $y$-coordinates provide a clear visualization of gait events, marking heel strikes and toe-offs for both left and right ankles. These events align with biomechanical expectations, validating the model’s temporal precision in identifying critical gait cycle points.
	
	The knee angle trajectories for both legs highlight variations in flexion and extension throughout the gait cycle. The close tracking of these angles over time offers valuable insights into joint coordination, with smooth transitions reflecting normal gait patterns. Deviations from these expected patterns may indicate potential gait abnormalities, which can be further analyzed using predictive uncertainty measures from the MoGP model.
	
	The phase duration plots show the distribution of stance and swing phases across multiple gait cycles. Both left and right legs exhibit consistent durations, indicating symmetrical gait. The balance between stance and swing phases is a key indicator of gait health, and discrepancies in their durations may signal underlying motor impairments. By breaking down these phases, the model enables deeper biomechanical analysis, helping clinicians assess abnormal phase transitions, such as prolonged stance or diminished swing phases.
	
	These results demonstrate the model's robustness in capturing multi-dimensional gait dynamics. By combining ankle positions, knee angles, and phase durations, the predictive outputs provide a comprehensive view of gait patterns, enabling accurate detection of abnormalities.
	
	\subsection{Analysis of Stance and Swing Phases in Gait Dynamics}
	
\begin{figure}[ht!]
		\centering
		\includegraphics[width=0.95\linewidth]{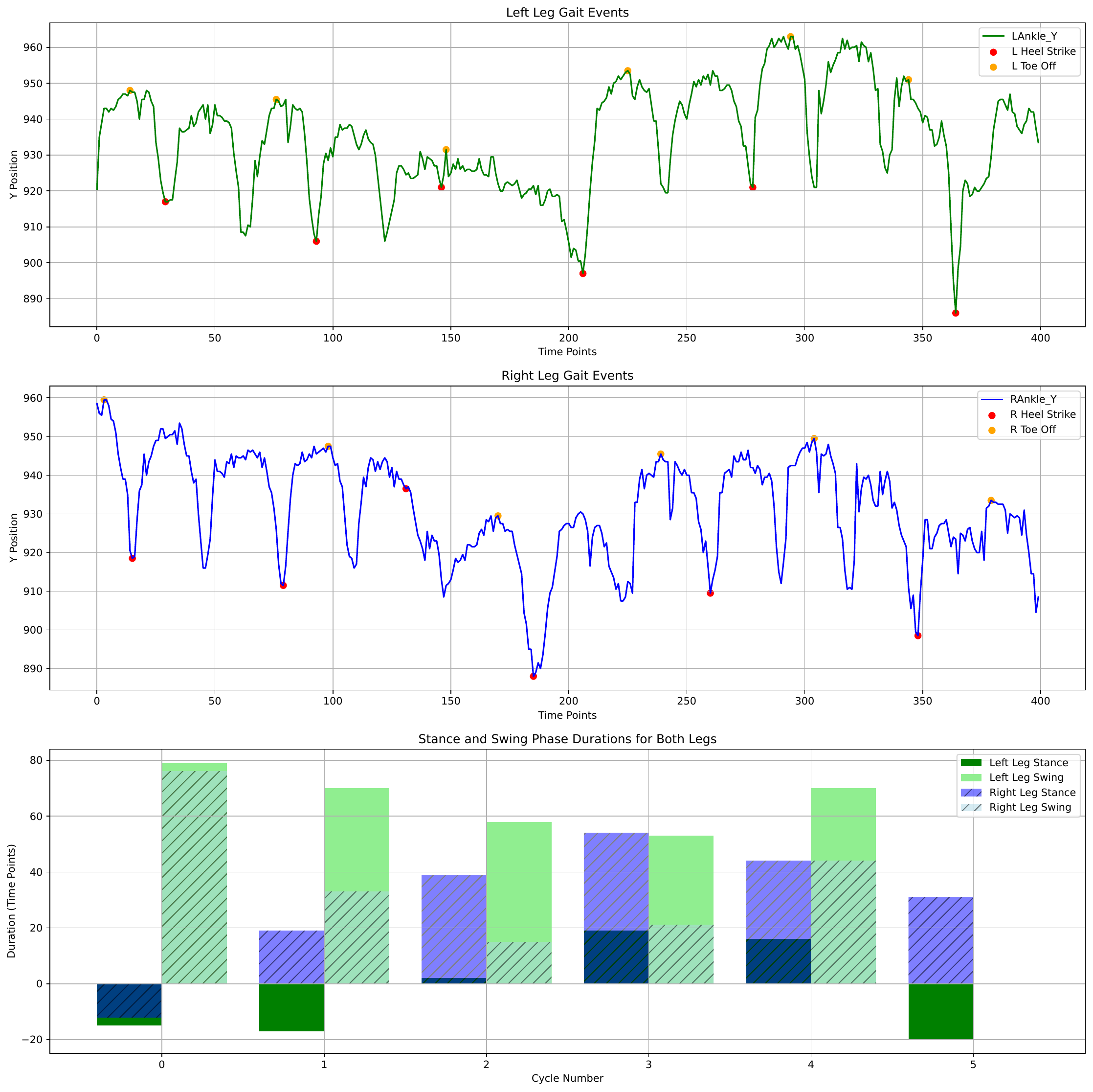}
		\caption{Stance and Swing Phase Durations with Gait Events. The top plots display the $y$-coordinates of the left and right ankles across time points, with markers indicating heel strikes and toe-offs, essential events in the gait cycle. The bottom plot illustrates the stance and swing phase durations for both legs over multiple gait cycles, highlighting the symmetry and consistency between the left and right legs. These visualizations offer detailed insights into gait dynamics, providing key metrics for assessing gait health and identifying potential abnormalities.}
		
		\label{fig:StanceSwingPhases}
	\end{figure}

	Figure \ref{fig:StanceSwingPhases} provides a detailed visualization of stance and swing phases for both legs, along with key gait events such as heel strikes and toe-offs. The ankle $y$-coordinates are plotted over time, with markers indicating these critical events. Their alignment with expected gait patterns verifies the model's precision in event detection.
	
	The phase duration plot illustrates stance and swing phase durations across multiple gait cycles. The results show consistent and symmetric durations between the left and right legs, an important indicator of healthy gait. The relative balance between stance and swing phases, typically reflecting efficient gait mechanics, is well captured, with no observable discrepancies across cycles. In cases of motor impairments, such as those caused by neurological conditions, these durations may vary, with prolonged stance or shortened swing phases serving as indicators of abnormal gait.
	
	These findings confirm the model's ability to accurately segment and quantify gait phases. The consistent prediction of gait events and symmetry in phase durations reinforce its reliability for clinical gait analysis. Such detailed phase assessments are crucial for evaluating gait health, detecting subtle abnormalities, and supporting early intervention in rehabilitation.
	
	\subsection{HMM-Based Gait Phase Segmentation and Abnormal Detection}
	
	\begin{figure}[ht!]
		\centering
		\includegraphics[width=0.95\linewidth]{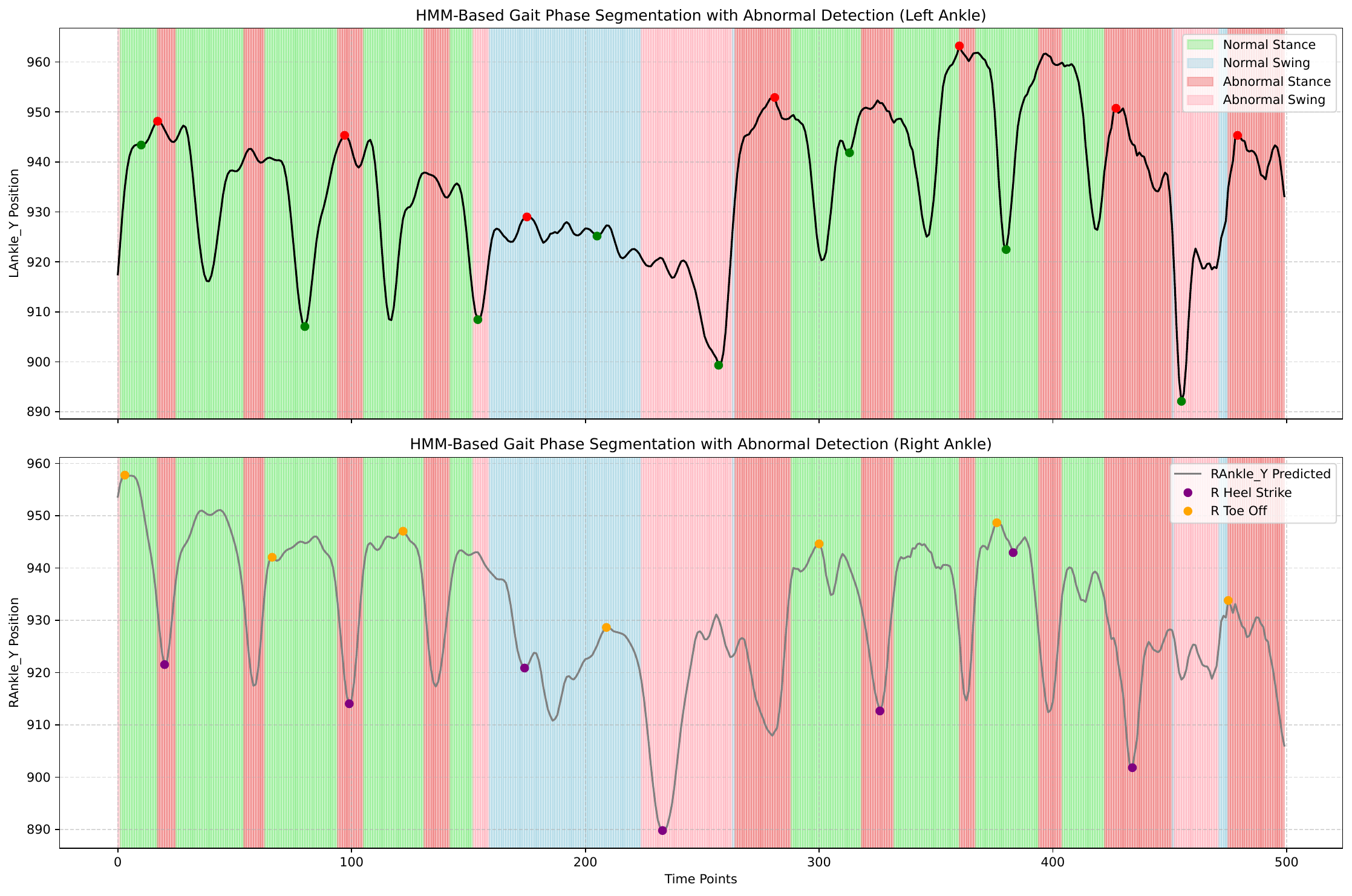}
		\caption{HMM-Based Gait Phase Segmentation with Abnormal Detection. The figure illustrates the segmentation of gait phases for the left and right ankles using a HMM. Four states are identified: Normal Stance, Normal Swing, Abnormal Stance, and Abnormal Swing, with key gait events such as heel strikes and toe-offs marked along the trajectories. This segmentation provides a detailed temporal view of gait dynamics, enabling the detection of deviations from normal patterns. }
		
		\label{fig:HMMPhaseSegmentation}
	\end{figure}
	
	Figure \ref{fig:HMMPhaseSegmentation} presents the HMM-based phase segmentation of left and right ankle trajectories, distinguishing between normal and abnormal gait phases. For each ankle, the model classifies gait into four states: Normal Stance, Normal Swing, Abnormal Stance, and Abnormal Swing, with transitions highlighted over time. The accurate identification of key events such as heel strikes and toe-offs demonstrates the HMM's ability to capture the cyclic structure of gait, providing detailed temporal segmentation aligned with biomechanical expectations.
	
	From a clinical perspective, the detection of abnormal phases offers valuable insights into motor impairments, allowing the identification of subtle gait deviations. The visualization of normal versus abnormal phases facilitates early detection of potential gait abnormalities, which is crucial for children with neurological conditions, such as those affected by perinatal asphyxia. This segmentation not only supports personalized clinical assessments but also informs rehabilitation strategies by highlighting specific gait deviations (e.g., abnormal stance durations), that may require targeted intervention. The robust tracking of these phases over multiple time points confirms the model's utility for long-term gait monitoring and intervention planning.
	
	\section{Conclusions}
	\label{sec:conclusions}
	
	This paper presents a method for modeling and analyzing gait dynamics by integrating a Multi-output Gaussian Process (MoGP) model with a Hidden Markov Model (HMM) for phase segmentation and abnormality detection. The MoGP model, which uses a customized composite kernel based on the sum of the periodic, SE and Mat\'ern $3/2$ kernels, accurately predicts joint trajectories and provides well-calibrated predictive intervals. The HMM segments gait into normal and abnormal phases, allowing precise temporal analysis of stance and swing cycles.
	
	The findings of this study demonstrate that the proposed MoGP-HMM framework effectively captures multidimensional gait patterns, enabling accurate identification of deviations associated with motor neurological disorders. By using a probabilistic framework, the model provides reliable phase tracking across multiple gait cycles, offering valuable insights for early diagnosis and rehabilitation planning. 
	
	The integration of a composite kernel, designed based on expert knowledge, enhances the model's ability to jointly capture individual joint dynamics and inter-joint correlations. This allows for robust estimation of key gait events, such as heel strikes and toe-offs, while maintaining consistent phase segmentation. The ability to quantify uncertainty strengthens the framework's clinical applicability, providing reliable decision-making in pathological gait assessment.
	
	From a clinical perspective, accurately predicting normal and abnormal gait phases enables early identification of motor impairments, such as those related to perinatal asphyxia. The strong alignment between predicted and observed gait trajectories validates the model's potential for real-time monitoring and longitudinal gait assessments. These results highlight the impact of probabilistic models in pediatric neurology, supporting the development of data-driven approaches that enhance clinical decision-making and facilitate personalized rehabilitation strategies.

	Future work will focus on developing a heterogeneous GP modeling framework that jointly estimates gait patterns and abnormal behavior by integrating multimodal data sources. This approach aims to enhance robustness by incorporating additional biomechanical, electrophysiological, and clinical markers, leading to a more comprehensive assessment of motor impairments.
	
	
	\bibliographystyle{IEEEtran}
	\bibliography{biblio}
	
\end{document}